\documentclass[twocolumn]{svjour3}          
\smartqed  
\usepackage[draft]{hyperref}
\usepackage{graphicx}
\usepackage{rotating}
\usepackage{lscape}
\usepackage{multirow}
\usepackage{natbib}

\usepackage{color}

\begin{document}

\title{Falls as anomalies? An experimental evaluation using smartphone accelerometer data
}

\titlerunning{Falls as anomalies?}        

\author{Daniela Micucci   \and
        Marco Mobilio \and
        Paolo Napoletano \and
        Francesco Tisato 
}



\institute{D. Micucci \and M. Mobilio \and Paolo Napoletano \and F. Tisato \at
              DISCo, University of Milano - Bicocca, Viale Sarca 336, 20126 Milan, Italy \\
              \email{daniela.micucci@unimib.it}           
}

\date{Received: date / Accepted: date}

\maketitle

\begin{abstract}
Life expectancy keeps growing and, among elderly people, accidental falls occur frequently. A system able to promptly detect falls would help in reducing the injuries that a fall could cause. Such a system should meet the needs of the people to which is designed, so that it is actually used. In particular, the system should be minimally invasive and inexpensive. Thanks to the fact that most of the smartphones embed accelerometers and powerful processing unit, they are good candidates both as data acquisition devices and as platforms to host fall detection systems. For this reason, in the last years several fall detection methods have been experimented on smartphone accelerometer data. Most of them have been tuned with simulated falls because, to date, datasets of real-world falls are not available.
This article evaluates the effectiveness of methods that detect falls as anomalies. To this end, we compared traditional approaches with anomaly detectors. In particular, we experienced the kNN and the SVM methods using both the one-class and two-classes configurations. The comparison involved three different collections of accelerometer data, and four different data representations. Empirical results demonstrated that, in most of the cases, falls are not required to design an effective fall detector.


\end{abstract}

\section{Introduction}
\label{intro}

Falls are a major health risk that impacts the quality of life of elderly people. When a fall occurs, a prompt notification would help in reducing the injuries that the fall could cause. 
An effective fall detection system should address the following requirements~\citep{abbate2012}: 1) automatic notification of occurred falls; 2) promptness in order to provide quick help; 3) reliability of the fall detection techniques; 4) communication capabilities in order to alert the caregivers; 5) usability in order to facilitating users' acceptance.

Several solutions have been proposed: some of them addressing the problem as a whole, and others focusing on one specific requirement. 
The contribution of this article is related to the reliability of the fall detection techniques.

Several factors characterize a fall detection technique: from the sensors used to acquire data, to the features extracted; from the algorithms used to detect falls, to the types of datasets used to train the algorithm. The approaches that have been proposed differ for the choices with respect to those factors.

For what concerns data acquisition, ambient sensors, wearable sensors, or a combination of the two, are the principal data sources used in these techniques \citep{mubashir_survey_2013,liming_chen_sensor-based_2012}. Many recent approaches investigate the possibility of using the sensors provided by smartphones~\citep{medrano2014,sposaro_ifall:_2009,abbate2012}, which are widespread and require almost no installation or set-up. Moreover, they do not introduce any additional cost, can be used in any place, and are accepted by end users because they are already part of their everyday life.

\begin{figure*}[tb]
  \centering
  \scriptsize
   \setlength{\tabcolsep}{2.1pt}
  \begin{tabular}{c}
  \includegraphics[width=.70\textwidth]{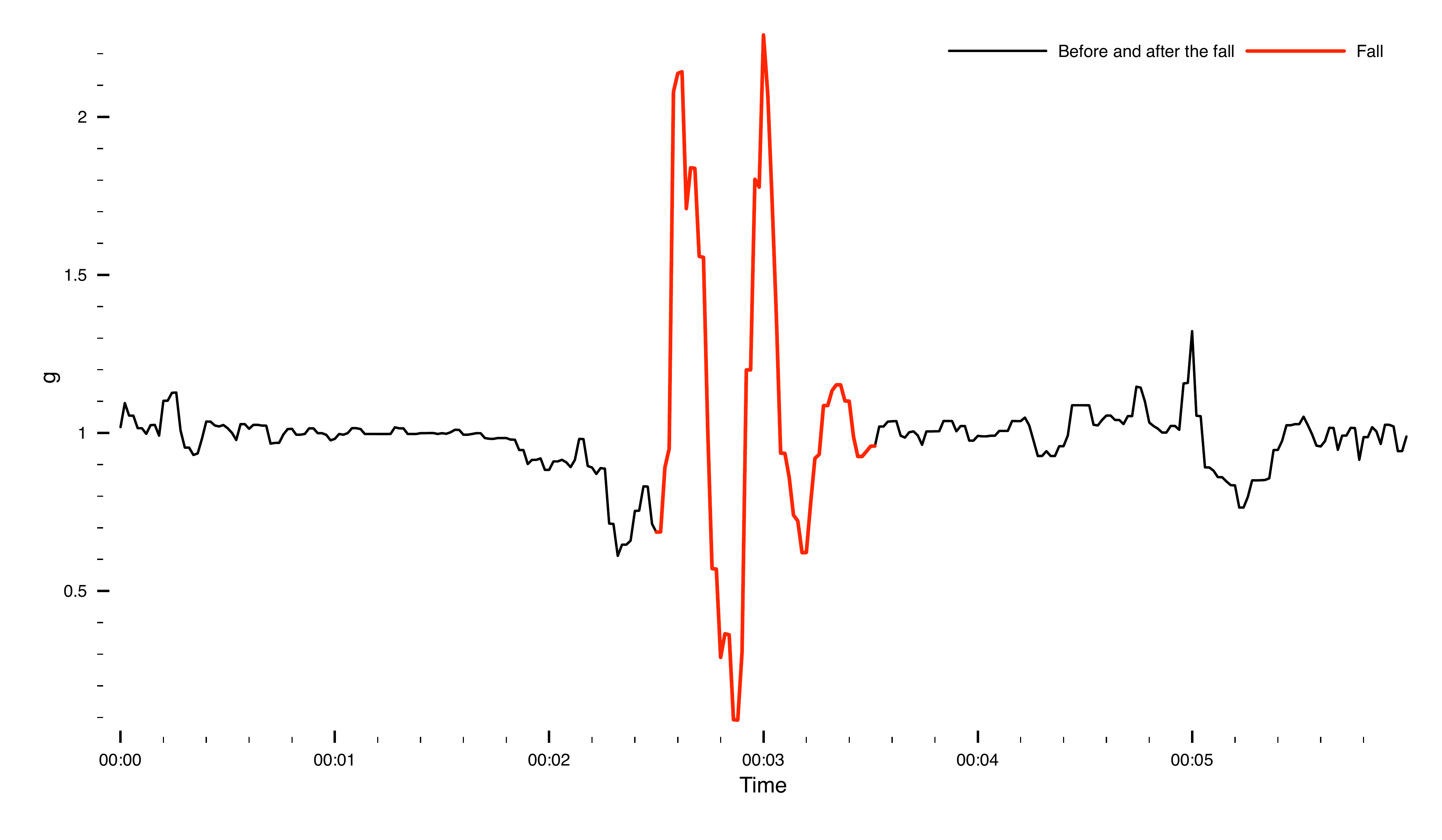} \\
  (a)\\
    \includegraphics[width=.70\textwidth]{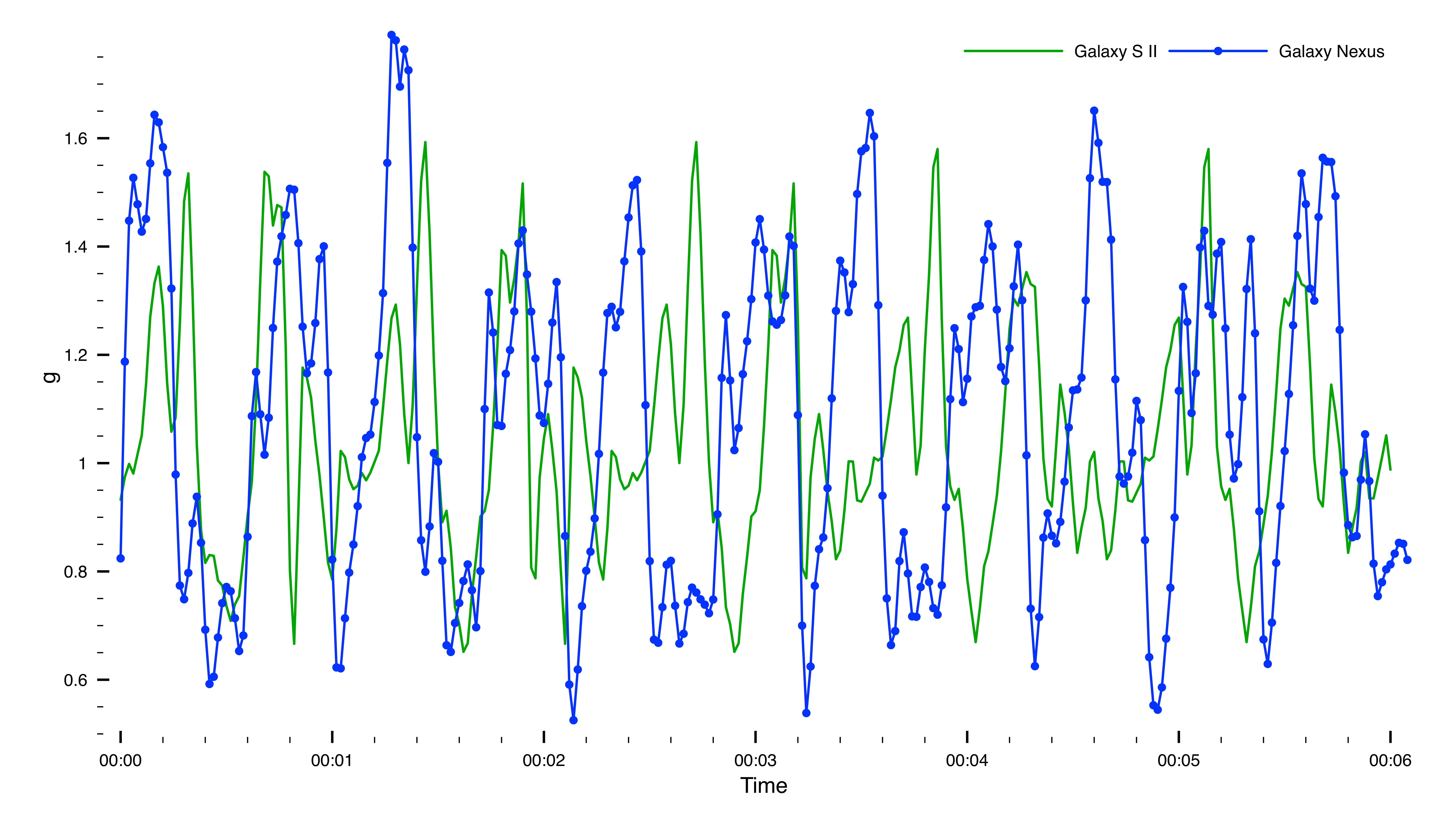} \\
  (b)\\
   \end{tabular}
  \caption{Examples of accelerometer data: (a) A fall as acquired by a smartphone. (b) A walking activity from two different smartphones performed by two different subjects.}
  \label{fig:accData}
\end{figure*}

The techniques also differ in the data type used to detect falls. Most popular fall detection techniques exploit accelerometer data as the main input to discriminate between falls and activities of daily living (ADL). Fig.~\ref{fig:accData}.a shows an example of accelerometer data representing a fall that are extracted from the dataset provided by~\citet{medrano2014}. In particular, the fall was recorded with a smartphone Galaxy mini. Fig.~\ref{fig:accData}.b illustrates the accelerometer data recorded by two sensors respectively placed on a Galaxy S II (from the dataset by~\citet{anguita2013public}) and a Galaxy Nexus (recorded by ourselves). These data capture the walk performed by two different subjects. It is possible to notice that the captured data share a general trend. This suggests the possibility of defining a method for the detection of falls that can be general and independent from the specific devices.

To verify the effectiveness of the method used by the technique to detect falls, data acquired by the sensors are arranged into labeled datasets containing both ADL and falls, usually simulated by volunteers. Often, datasets are elaborated in order to obtain features: from simple raw data to more complex indicator (e.g., magnitude and Fourier transform) whose processing requires time and computational resources. Methods can be principally divided into two main categories: domain knowledge- and machine learning techniques-based\linebreak \citep{mirchevska_combining_2014}. The approaches currently proposed, regardless of their classification, have in common the fact that they require a set of falls in their training phase. Unfortunately, human simulations are significantly different from real-world falls~\citep{klenk_comparison_2011}, and this could make those fall detection techniques not feasible for real-world applications.

For this reason,~\citet{medrano2014} experimented the use of a machine learning technique based on one-class classifier that has only been trained on ADL to detect falls as anomalies with respect to ADL. In particular, their experimentation was conducted with a k-Nearest Neighbour (kNN) classifier. As data representation they used the magnitude that does not require an huge amount of resources to be calculated. 
\citet{medrano2014} experimented on a publicly available dataset containing both ADL and falls 
 simulated by several human subjects and recorded by the same device. Moreover,~\citet{medrano2014} also experienced a two-classes Support Vector Machine (SVM) on the same dataset. SVM has produced slightly better results with respect to the one-class kNN. Thus, they concluded that anomaly detectors are infeasible in detecting falls.
Finally, in the article they explicitly state that data are acquired by accelerometers mounted on smartphones. This suggests that it was taken into consideration the idea of running the analysed methods on smartphones. From our point of view, a smartphone hardly support the execution of a SVM ensuring good performance.
Indeed, as~\citet{Mazhelis06} states, SVMs feature very high computational requirements for training. Since the final aim is also to provide a continuous learning system, the high complexity of the training phase is critical when the deployment occurs on a mobile device with limited computational power and, most of all, with limited power resources. Indeed, energy consumption is today one of the main issue in mobile computing, especially when dealing with physical sensors (\citet{Pejovic:2015}).

The aim of our work is to evaluate the effectiveness of methods that detect falls as anomalies with respect to traditional approaches that use two-classes classifiers to distinguish between falls and ADL. We compared anomaly detectors based on one-class kNN and SVM with traditional detectors based on two-classes kNN and SVM. We considered four different data representations calculated from accelerometer data acquired by smartphones: \emph{raw data}, \emph{magnitude}, \emph{accelerometer features}, and \emph{local temporal patterns}.  We evaluated the classifiers with respect to the variations of acquisition conditions: different sensors, different human subjects, different sensor positions. All the experiments have been conducted on two publicly available datasets~\citep{medrano2014,anguita2013public} of accelerometer data acquired by smartphones. Evaluation metrics, such as area under the curve (AUC), sensitivity and specificity, confirmed, in most of the cases, that anomaly detection techniques are quite robust against variations of acquisition conditions. 

The rest of the paper is organised as follows: Section \ref{sec:related_work} introduces the motivations of our work and discusses related work; Section \ref{sec:ExperimentDesign} outlines the experiment design; Section \ref{sec:results} presents the results of the experimentations; Section \ref{sec:discussion} discusses the achieved results; finally Section \ref{sec:future} provides some details about the future directions. 
\section{Motivation and Related Work}
\label{sec:related_work}

In the near future the number of elderly people is expected to grow. Indeed, the World Population Ageing Report states that the global share of elderly people (aged 60 years or over) will reach more than 21\% by 2050 (more than 2 billion people)~\citep{WPA2013:website}. Ageing results from the demographic transition, a process where reductions in mortality are followed by reductions in fertility~\citep{WPA2013:website,carone_can_2006}. The increasing trend of life expectancy has been directly proportional to the increase in disability~\citep{karmarkar_prescription_2009}. Thus, oldest people represent the greatest challenge in providing health-related services and identifying ways to assist them in maintaining independence~\citep{mann_aging_2004}. Indeed, the 31.2\% of people aged 80 to 84, and 49.5 percent of those over age 85, require assistance with everyday activities~\citep{older_2012}. This increment results in a growing need for supports (human or technological) that enable the older population to perform daily activities~\citep{us_census_bureau_international_2013}.

Intensive research efforts have been and are still focused on the identification of solutions that from one side automatically assist elderly people in performing daily activities and, on the other side, promptly detect anomalous situations related to diseases or to situations purely related to the old age, such as the worsening of the mild cognitive impairment~\citep{acampora_survey_2013}, the prompt identification of conditions favorable to heart failures~\citep{deshmukh_wearable_2015}, and the prompt detection of falls~\citep{mubashir_survey_2013}.

Falls are a major health risk that impacts the quality of life of elderly people. Among elderly people, accidental falls occur frequently: the 30\% of the over 65 population falls at least once per year; the proportion increases rapidly with age~\citep{tromp2001}. Moreover, fallers who are not able to get up more likely require hospitalization or, even worse, die~\citep{tinetti1993}. Thus, several approaches have been proposed to prompt detect falls. They mainly differ with respect to (i) the sensors used to acquire data, (ii) the data representation (features) used by the method, and (iii) the method used to detect falls.

Table~\ref{relatedWork} summarizes the analysis performed on a set of significative approaches. The table has been specifically designed to highlight the characteristic features of each approach in terms of (i) sensors, (ii) data representation and (iii) methods. In particular, the first two columns show the method and the training set configuration respectively. The third column states whether the approach requires a set of falls to train the algorithm. The fourth column lists the set of features used to infer a fall. Finally, the fifth and sixth columns respectively specify the type of wearable sensor used to sense data (ad hoc solutions or smartphone's sensors) and the involved sensors.

\begin{table*}[htbp]
\scriptsize
\caption{Related work}
\label{relatedWork}
\begin{center}
\begin{tabular}{llclclp{0.5in}}

\hline
\textbf{Approach}
& \textbf{Method}
& \begin{tabular}[c]{@{}c@{}}\textbf{Falls}  \\ \textbf{needed?}\end{tabular}
& \textbf{Features}
& \begin{tabular}[c]{@{}c@{}}\textbf{Smartphone} \\ \textbf{Ad-hoc}\end{tabular} & \textbf{Sensors} \\
\hline \hline

\citet{medrano2014}
& K-means+NN
& no
&- Magnitude
& Smartphone
&- Triaxial accelerometer \\
\hline

\citet{tolkiehn2011}
& Threshold based
& yes
& \begin{tabular}[c]{@{}l@{}}
    - Magnitude of standard \\\hspace{1 mm} deviation per axis\\
    - Std of the magnitude\\
    - Ratio of the polar angle\\
    - Delta of two consecutive \\\hspace{1 mm} polar angles\\
    - Barometric pressure
   \end{tabular}
& Ad-hoc
& \begin{tabular}[c]{@{}l@{}}
    - Triaxial accelerometer \\
    - Barometric pressure
    \end{tabular}
\\ \hline

\citet{wang2014}
& Threshold based
& yes
& \begin{tabular}[c]{@{}l@{}}
    - Signal magnitude vector \\
    - Hearth rate value\\
    - Trunk angle
    \end{tabular}
& Ad-hoc
& \begin{tabular}[c]{@{}l@{}}
    - Triaxial accelerometer \\
    - Hearth rate monitor
    \end{tabular}
\\ \hline

\citet{bourke2007}
& Threshold based
& yes
&- Magnitude & Ad-hoc
& \begin{tabular}[c]{@{}l@{}}
    - Dual-axis accelerometers\\\hspace{1 mm} placed orthogonally
    \end{tabular}
\\ \hline

\citet{li2009}
& Threshold based
& yes
& \begin{tabular}[c]{@{}l@{}}
    - Magnitude of acceleration \\
    - Magnitude of angular\\\hspace{1 mm} velocity
    \end{tabular}
& Ad-hoc
& \begin{tabular}[c]{@{}l@{}}
    - Triaxial accelerometer\\
    - Triaxial gyroscope
    \end{tabular}
\\ \hline

\citet{zhang2006}
& One-class SVM
& yes
& \begin{tabular}[c]{@{}l@{}}
    - Time of free fall \\
    - Variance of acceleration \\\hspace{1 mm} during free fall\\
    - Time of reverse impact \\
    - Mean and variance of \\\hspace{1 mm} acceleration during \\\hspace{1 mm} reverse impact\\
    \end{tabular}
& Ad-hoc
&- Triaxial accelerometer                                                                                                    \\ \hline

\citet{chen2006}
& Threshold based
& yes
&- Magnitude                                                                                                                                                                                                           & Ad-hoc
& \begin{tabular}[c]{@{}l@{}}
    - Dual-axis accelerometers \\\hspace{1 mm} placed orthogonally
    \end{tabular}
\\ \hline

\citet{nyan2008}
& Threshold based
& yes
& \begin{tabular}[c]{@{}l@{}}
    - Correlation coefficient \\\hspace{1 mm} between thigh and waist \\\hspace{1 mm} deviation from vertical axis\\
    - Correlation coefficient \\\hspace{1 mm} between angular velocity \\\hspace{1 mm} and reference template
    \end{tabular}
& Ad-hoc
& \begin{tabular}[c]{@{}l@{}}
    - Triaxial accelerometer\\
    - Two-axis gyroscope
\end{tabular}\\
\hline

\citet{abbate2012}
& Threshold based
& yes
& \begin{tabular}[c]{@{}l@{}}
    - Magnitude\\
    - Time of inactivity\\
    - Peak time\\
    - Impact start\\
    - Impact end
    \end{tabular}
& \begin{tabular}[c]{@{}c@{}}
    Smartphone \\
    Ad-hoc
    \end{tabular}
&- Triaxial accelerometer \\
\hline

\citet{ge2008}
& Threshold based
& yes
& \begin{tabular}[c]{@{}l@{}}
    - Inertial frame vertical\\ \hspace{1 mm} acceleration\\
    - Inertial frame vertical \\\hspace{1 mm} velocity\\
    - Time of free fall
    \end{tabular}
& Ad-hoc
& \begin{tabular}[c]{@{}l@{}}
    - Triaxial accelerometer\\
    - Two-axis gyroscope
    \end{tabular}\\
\hline

\citet{mellone2012}
& Threshold based
& yes
& \begin{tabular}[c]{@{}l@{}}
    - Acceleration sum vector\\
    - Vertical axis orientation
    \end{tabular}
& Smartphone
&- Triaxial accelerometer
\\ \hline

\citet{rabah2012}
& Threshold based
& yes
& \begin{tabular}[c]{@{}l@{}}
    - Magnitude\\
    - Orientation
    \end{tabular}
& Ad-hoc
&- Triaxial accelerometer
\\ \hline

\citet{7061032}
& SVM
& yes
& \begin{tabular}[c]{@{}l@{}}
    - Range of angular velocity\\
    - Range of acceleration
    \end{tabular}
& Ad-hoc
& \begin{tabular}[c]{@{}l@{}}
    - Triaxial accelerometer\\
    - Two-axis gyroscope\\
    - Single axe gyroscope
\end{tabular}
\\ \hline

\citet{zhuang2015fall}
& SVM
& yes
& \begin{tabular}[c]{@{}l@{}}
    - Magnitude\\
    - Ascending coefficient\\
    - Descending coefficient
    \end{tabular}
& Ad-hoc
&- Triaxial accelerometer
\\ \hline

\citet{sposaro_ifall:_2009}
& Threshold based
& yes
& \begin{tabular}[c]{@{}l@{}}
    - Magnitude\\
    - Angle of body
    \end{tabular}
& Smartphone
&- Triaxial accelerometer\\
\hline

\citet{dai_mobile_2010}
& Threshold based
& yes
& \begin{tabular}[c]{@{}l@{}}
    - Magnitude\\
    - Ad-hoc feature\\
    - Shape Context\\
    - Hausdorff distance
    \end{tabular}
& \begin{tabular}[c]{@{}c@{}}
    Smartphone \\
    Ad-hoc
    \end{tabular}
& \begin{tabular}[c]{@{}l@{}}
    - Triaxial accelerometer\\
    - Magnetometer\end{tabular} \\
\hline

\citet{lee_detection_2011}
& Threshold based
& yes
& \begin{tabular}[c]{@{}l@{}}
    - Magnitude\\
    - Mean of magnitude
    \end{tabular}
& Smartphone
& - Triaxial accelerometer\\
\hline

\citet{albert_fall_2012}
& SVM
& yes
& \begin{tabular}[c]{@{}l@{}}
    - Moment (mean, abs(mean),\\ \hspace{1 mm} std, skew, kurtosis\\
    - Moments of the difference\\ \hspace{1 mm} between successive samples\\
    - Smoothed root mean squares\\
    - Extremes (min, max,\\ \hspace{1 mm} abs(min), abs(max))\\
    - Histogram\\
    - Fourier components\\
    - Mean magnitude, mean of\\ \hspace{1 mm} cross products (xy, xz, yz),\\ \hspace{1 mm} abs(mean of cross products)\\
    \end{tabular}
& Smartphone
&- Triaxial accelerometer\\
\hline

\citet{fang_developing_2012}
& Threshold based
& yes
& \begin{tabular}[c]{@{}l@{}}
    - Magnitude\\
    - Vertical acceleration value\\
    \end{tabular}
& Smartphone
& \begin{tabular}[c]{@{}l@{}}
    - Triaxial accelerometer\\
    - Gyroscope\\

\end{tabular}\\
\hline
\end{tabular}
\end{center}
\end{table*}

Table~\ref{relatedWork} aims at providing  an idea of how many different approaches are proposed. Most of the approaches rely on data coming from ad-hoc wearable sensing devices, only a few on smartphone's sensors. The mainly used sensors are accelerometers. The approaches use features that are very different each other, some of them very complex in terms of computation. Half of the approaches is based on thresholds-based techniques and the other half on machine learning techniques. Finally, most of the approaches are based on methods that require a set of fall to train the underlying algorithm.

Other approaches not outlined in Table \ref{relatedWork} can be found in the many surveys dedicated to the fall detection (e.g., ~\citet{mubashir_survey_2013,mohamed_fall_2014,hijaz_survey_2010}). Among the others,~\citet{bagala_evaluation_2012} is particularly interesting because it compares the most popular techniques for the identification of falls based on accelerometer data.

\subsection{Sensors and Data Representation}
Fall detection methods rely on data acquired by sensing devices. Images, accelerometer data, audio, angular velocities are only a few examples of data. Data are captured by environmental or wearable sensors or by a combination of both~\citep{mubashir_survey_2013,liming_chen_sensor-based_2012}.
Ambient sensors introduce many issues such as privacy, installation costs, and invasiveness. Moreover, a person can fall everywhere. Thus, wearable senors are more indicated for the specific application domain. Under the umbrella of wearable sensors fall ad-hoc solutions and smartphones' sensors. Ad-hoc solutions generally include a microcontroller and a set of attached sensors. Such artifacts are then placed in specific area of the body (e.g., wrist, arm, ankle). Thus, they require an explicit acceptance by the elderly people. On the opposite, smartphones are generally present in everyday life. Therefore, the use of smartphones do not require changes in daily habits and do not involve additional costs.

Despite the type of wearable device, most of the approaches use accelerometers, a few accelerometers with gyroscopes. For this reason, our experimentation has considered accelerometers only.

As regards features, Table~\ref{relatedWork} shows how the various approaches use features of different nature. Therefore, there not exists a common trend. The unique feature that is found with greater frequency is magnitude.

It is possible to notice that some of the used features are generic such as the magnitude, the energy, and the standard deviation. Others are specifically related to the application domain, such as the time of free fall, the time of reverse impact, and the time of inactivity.

Some of them are performing (such as, magnitude and Fourier transform), but require high processing times and/or considerable computing resources with respect to the application domain and, in case of smartphone, to the device on which the features will be calculated. Indeed, timeliness and lightness in the computation are crucial factors so that those features can be used with effectiveness on smartphones.

\subsection{Methods}
Methods can be divided into domain knowledge- and machine learning techniques-based~\citep{mirchevska_combining_2014}: the former usually apply heuristics, while the latter usually rely on the definition of classifiers able to detect falls. From our perspective, regardless of the type, what differentiates the techniques is the need for data representing falls in the data set used to train the algorithm. Most of the proposed approaches require falls in the training data set in order to properly configure their method. Falls are mostly realized relying on volunteers that are asked to perform daily activities (such as, sitting, walking, and so on) and to simulate falls.

Even if the achieved results by those approaches are very promising, it is quite difficult to generalize the results because almost always the experimentation is limited to one ad-hoc data set only. In addition, as stated by~\citet{klenk_comparison_2011}, simulated falls significantly differ from real-world falls. Thus, having simulated falls in the training dataset could lead to realize classifiers that may show different behaviours with real-world falls.

For the above considerations, a method based on the detection of anomalies with respect to ADL may be a better solution for this kind of application domain. Fall detection is not the only case in which the detection of anomalies is the better choice in designing a classifier. There are many other situations in which real-world data are very difficult to achieve: imagine a system able to infer terrorist attacks. Real world training sets are very rare or even not existing. From the related work analysis, only~\citet{medrano2014} have assessed the robustness of one-class classifiers trained with a set of ADL. Indeed, \citet{medrano2014} agree with us stating that ``traditional approaches to this problem suffer from a high false positive rate, particularly, when the collected sensor data are biased toward normal data while the abnormal events are rare''. 

\citet{medrano2014} also experimented a two-classes SVM (Support Vector Machine). They concluded that SVM allows to obtain best results with respect to one-class kNN classifier in detecting anomalies. Although the goodness of the results,  the training of a SVM would be more expensive in terms of computation resources than the training process of a kNN-based  one-class classifier.

\section{Experiment Design}
\label{sec:ExperimentDesign}

In this article we focus on methods that detect falls exploiting smartphone accelerometer data. In particular, we evaluated the robustness of anomaly detectors compared to traditional detectors that, in turn, are tuned with fall instances. To this end, we designed several experimental setups by varying both \emph{materials and methods}:

\begin{itemize}
\item \textbf{Data}: we have created three different collections of smartphone accelerometer data by mixing the data of two publicly available smartphone accelerometer data~(\citet{medrano2014} and~\citet{anguita2013public}) that have been recorded by different devices with different setups. We have created two sets of these collections selecting different sizes of time window of the accelerometer data. Experimenting on these collections permits to assess the robustness with respect to changes in acquisition conditions.
\item \textbf{Feature vectors}: we experimented four different feature vectors ranging from the most simple to the most complex ones: \emph{raw data}, \emph{magnitude}, \emph{accelerometer features}, \emph{local temporal patterns}. Assessing the goodness of feature vectors is very meaningful especially in a mobile and real time environment where the computational capacity may be limited.
\item \textbf{Classification schema}: we compared two different classification schemas based on the k-Nearest Neighbour (kNN) classifier~\citep{KNN1951}: \emph{one-class} (corresponding to the anomaly detector) versus \emph{two-classes}. Moreover, we also compared the results achieved with the \emph{one-class} and the \emph{two-classes} classification schemas based on the Support Vector Machines (SVM) classifier~\citep{SVM1995}.
\end{itemize}

\subsection{Publicly Available Datasets}
One of the considered datasets contains both Activities of Daily Living (ADL) and falls performed by ten participants, 7 males and 3 females, ranging from 20 to 42 years old~\citep{medrano2014}. The ADL set has been recorded during real-life conditions: participants carried a smartphone in their pocket for at least one week to record everyday behaviour. On average, about
800 ADL instances were collected from each subject during this period. Participants simulated eight different types of falls:
\emph{forward falls}, \emph{backward falls}, \emph{left and right-lateral falls}, \emph{syncope}, \emph{sitting on empty chair}, \emph{falls using compensation strategies to prevent the impact}, and \emph{falls with contact to an obstacle before
hitting the ground}. Participants wore a smartphone in both their two pockets (left and right) and performed the falls on a soft mattress in a laboratory
environment. They repeated each fall three times for a total of 24 fall simulations. The dataset contains 503 falls\footnote{The authors declared that due to technical issues some falls had to be repeated in a few cases, so the number is higher than
$24\times2\times10$.} and 7816 ADL.

Accelerations have been recorded through the
built-in triaxial accelerometer of a Samsung Galaxy Mini phone
running the Android operating system version 2.2. The sampling
rate was not stable, with a value of about $45\pm12$ Hz. During the daily life monitoring,
whenever a peak in the acceleration magnitude was detected to be
higher than 1.5 $g$ ($g$ = gravity acceleration), a new data instance was recorded. Each data instance included information in a time window of 6 $s$ around the peak.
During each fall simulation, a 6 $s$ width time window
around the highest peak has been recorded. Afterwards, the offset error of each axis was removed and an interpolation was performed to get a sample every 2 $ms$ (50 Hz). We will refer to this set of data as \emph{dataset1}.

The other dataset considered contains only ADL performed by a group of 30 volunteers with ages ranging from 19 to 48 years~\citep{anguita2013public}. Each person was instructed to follow a protocol of 6 activities: \emph{standing}, \emph{sitting}, \emph{laying down}, \emph{walking}, \emph{walking downstairs}, and \emph{upstairs}. Each subject performed the protocol twice while wearing a Samsung Galaxy S II smartphone: on the first trial the smartphone was fixed on the left side of the belt, and on the second it was placed by the user himself as preferred. The tasks were performed in laboratory conditions but volunteers were asked to perform freely the sequence of activities for a more naturalistic dataset. The accelerometer signals were pre-processed by applying noise filters and then sampled in fixed-width sliding windows of 2.56 $s$ and 50\% overlap, thus obtaining 128 readings/window. The total number of accelerometer instances is 10299. We will refer to this set of data as \emph{dataset2}.

For the analysis presented in this paper, we considered two different sub-windows of the accelerometer patterns taken around  the peak. More in details, we considered two sub-windows of:
\begin{itemize}
\item  2.56 $s$ corresponding to a vector 128 samples;
\item 1 $s$ corresponding to a vector of 51 values.
\end{itemize}

\subsection{Data Collections Description}
As discussed before, we have created three different collections of smartphone accelerometer data by mixing the data of two publicly available smartphone accelerometer data (\citet{medrano2014} and \citet{anguita2013public}). For the evaluation we used a 10-fold cross-validation approach. The three data collections are then composed of 10 folds, each containing 90\% of training data and 10\% of test data. More in details:
\begin{itemize}
\item \emph{Collection 1}. ADL: 7035 training and 781 test data. FALL: 453 training and 50 test data.  Both ADL and FALL data have been taken from the \emph{dataset1};
\item \emph{Collection 2}. ADL: 7035 training and 781 test data. Half of ADL data have been randomly taken from the \emph{dataset1} and half from  the \emph{dataset2};  FALL: 453 training data and 50 test data.  All the FALL data have been extracted from the  \emph{dataset1};
\item \emph{Collection 3}: ADL: 9270 training data and 1029 test data. All the ADL data have been taken from the \emph{dataset2}; 453 FALL training data and 50 FALL test data.  All the FALL data have been extracted from the  \emph{dataset1}.
\end{itemize}

\subsection{Feature vectors}

As discussed before, we considered four different feature vectors extracted from two windows with different size: 2.56 $s$ corresponding to a vector 128 samples and 1 $s$ corresponding to a vector of 51 values. More in details we considered:

\noindent \textbf{Raw data}. This is the simplest representation. Each data instance is composed of the concatenation of the three accelerometer data ($x,y,z$), one for each direction. We obtain a final feature vector of size 384 for the case of 128 samples and 153 for the case of 51 samples.

\noindent \textbf{Magnitude}. This vector of features has been obtained from the three accelerometer data ($x,y,z$) as follows: $$M=\sqrt{x^2+y^2+z^2}.$$ We obtained a vector of size 128 and another of size 51.

\noindent \textbf{Accelerometer features}. These feature vectors have been obtained by concatenating four different features for each direction: \emph{mean of the acceleration values}, \emph{standard deviation of the acceleration values}, \emph{energy of the acceleration}, and  \emph{correlation of the acceleration}. The dimension of the final feature vector is of size 12. The \emph{energy of the acceleration}  is calculated as follows: $$Energy=\sqrt{\frac{\sum_{i=1}^{N}|a_{fft_{i}}|^2}{N}},$$ where N is the number of samples of the acceleration patterns and $a_{fft_{i}}$ are the fast Fourier transform  components of the input patterns. The \emph{correlation of the acceleration} is calculated between couple of directions: $x$ versus $y$, $x$ versus $z$, etc.

\noindent \textbf{Local Temporal Patterns}. This feature is the most complex representation. The feature vector is composed of the concatenation of acceleration patterns achieved by comparing the magnitude of each sample ($M_{s}$) with the several boosted magnitude values of $N$ neighbour samples ($M_{i}$). The boosted magnitude values corresponding to a given neighbour $i$ are achieved by increasing the original magnitude by an increasing decimal factor: $$M_{i}^{n} = n+M_{i},$$ with $n$ ranging from 0 to $M_{max}$. The value $M_{max}$ corresponds to the nearest decimal value of maximum magnitude. For each sample we have $M_{max}+1$ comparisons as results of the following inequality:  $$M_{s} > M_{i}^{n}.$$ The result of each comparison is represented as a binary vector map of size $N$ with 1 indicating if the above inequality is satisfied and 0 if not.  All the comparison maps are then summed in order to obtain a single vector map of size $N$. The number of neighbours has been set to $N=6$. The final feature vector is obtained by concatenating all the maps achieved for each sample thus obtaining $6\times 51 = 306$ and $6\times 128 = 768$.

\subsection{Methods and Their Evaluation}
\label{subsec:Methods}

We have considered the {one-class} k-Nearest Neighbour (kNN) classifier and the {one-class} SVM classifier as anomaly detection techniques. These classifiers have been trained only with ADL instances and tested with both ADL and FALL instances. Given a test instance, if the anomaly score is higher than a given threshold, the new instance is classified as an anomaly/fall, otherwise is classified as an ADL. By varying the threshold, the receiver
operating characteristic curve ($ROC$) and the area under the $ROC$ curve ($AUC$) can be obtained. We calculated also a specific value of sensitivity ($SE = \frac{TruePositives}{Positives}$) and specificity ($SP = \frac{TrueNegatives}{Negatives}$). These values have been obtained by selecting the point that maximised their geometric mean $\sqrt{SE\cdot SP}$, in a $ROC$ curve averaged over the cross-validation results.

We compared the anomaly detectors with a \emph{two-classes} kNN and a \emph{two-classes} radial basis SVM. These classifiers have been trained and tested with both instances of ADL and FALL. We converted the distances achieved by the kNN into scores ranging from 0 to 1. By thresholding the scores we drawn the ROC curve.

All the algorithms have been implemented in Matlab and tested with a PC equipped with the Ubuntu 14.10 distribution. Regarding the two variants of kNN for each fold of the 10-cross validation we have performed an inner 10-cross validation for choosing the best number of neighbours $k$. We  experimented 10 values of $k$ ranging from 1 to 10, and we used the Euclidean distance as distance measure. Regarding the SVM classifier, we used the built-in Matlab package. For each fold of the 10-cross validation we have performed an inner 10-cross validation for choosing the best regularisation and kernel parameters. The Matlab implementation of the SVM allows to achieve scores as outputs along with decisions. By thresholding these scores we obtain the corresponding  $ROC$ curve.

In order to make the results reproducible,  the data collections as well as the Matlab code used for the experiments are available on the authors' website\footnote{http://www.sal.disco.unimib.it/research/ambient-assisted-living/}.

\section{Results} \label{sec:results}
In the tables~\ref{table:results1},~\ref{table:summary-1-128},~\ref{table:summary-2-128},~\ref{table:summary-3-128} we report the results achieved on the three collections by all the classification schemas and feature vectors in the case of accelerometer data instances made of 128 samples. It is quite evident that the two-class SVM performs better than the other classification schemas especially over the first two collections. It is also quite clear that \emph{raw data} and \emph{energy} feature vectors perform better than the others with the  \emph{raw data} being the best. As can be noticed, the \emph{one-class} and \emph{two-classes} kNN achieve close performance in most cases. 

It is clear that the \emph{raw data} (the simplest feature vector) is one of the best feature vectors independently of collection and classification schema (see Table~\ref{table:summary-1-128} and Table~\ref{table:summary-2-128}). This result is a quite new to scientific community since the most used features are usually more complex, such as \emph{magnitude} and \emph{energy features}.

The results achieved on the third collection by all the feature vectors and classification schemas suggest that this collection is not challenging. In fact, this collection has been made by using the ADL from \emph{dataset2} and the FALL from the \emph{dataset1}. The results obtained on this collection depends on the fact that the experimentation protocol of the underlying datasets is quite different. In fact, in the case of the \emph{dataset1} the ADL were obtained by recording at least one week of daily life, while in the case of the \emph{dataset2}, each person was instructed to perform 6 ADL in a laboratory environment. The difference between the ADL data in the two datasets makes the classification problem too easy.

In the tables~\ref{table:results2},~\ref{table:summary-1-51},~\ref{table:summary-2-51},~\ref{table:summary-3-51} we report the results achieved on the three collections by all the classification schemas and feature vectors in the case of accelerometer data made of 51 samples. Here, the \emph{one-class} and \emph{two-classes} classifiers achieve close performance in every cases. It should be noticed that the performance achieved in the case of the 51 samples by all the kNN based solutions is better than the case of 128 samples.  Moreover, even in this case, raw data demonstrated to work better than or comparable with more complex feature vectors.

\begin{figure*}[tb]
  \centering
  \scriptsize
   \setlength{\tabcolsep}{2.1pt}
  \begin{tabular}{cc}
  \includegraphics[width=0.52\textwidth]{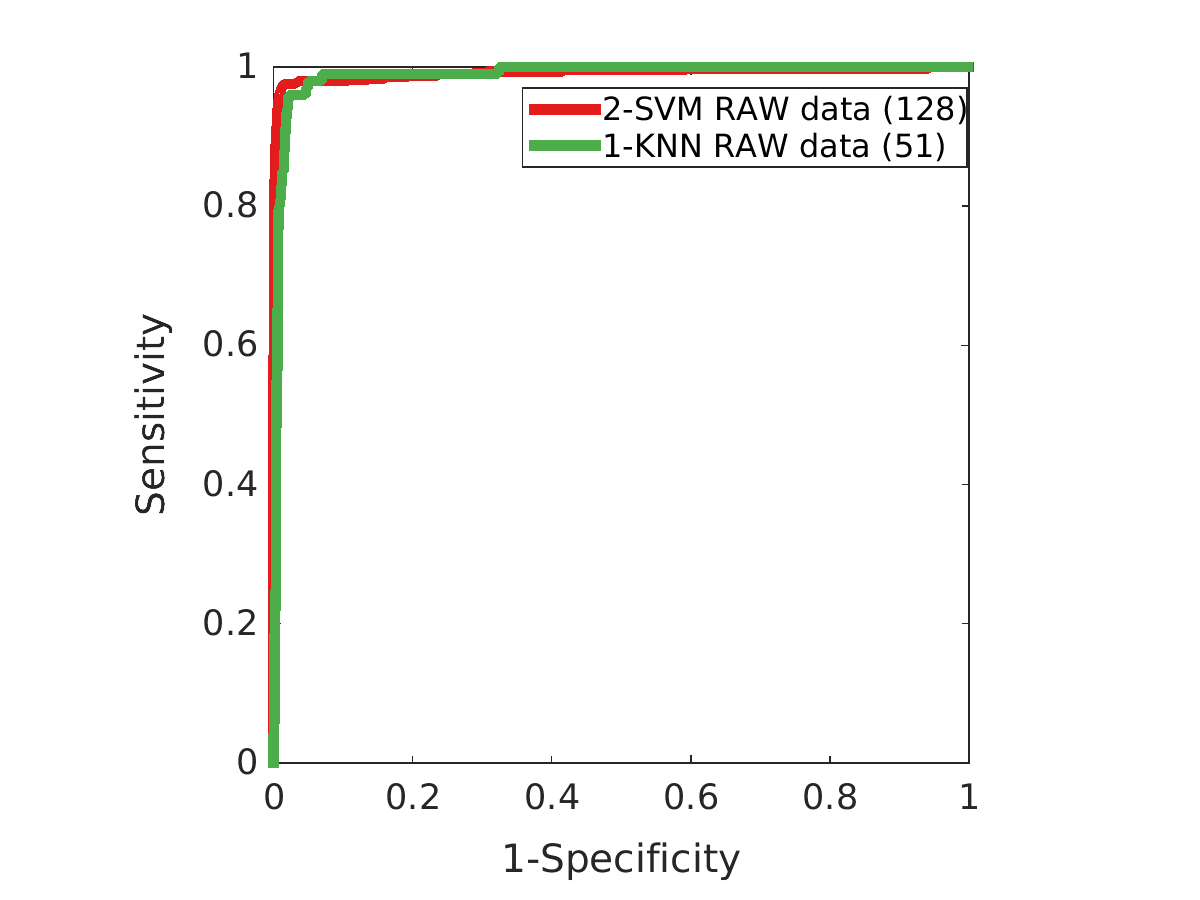} &
  \includegraphics[width=0.52\textwidth]{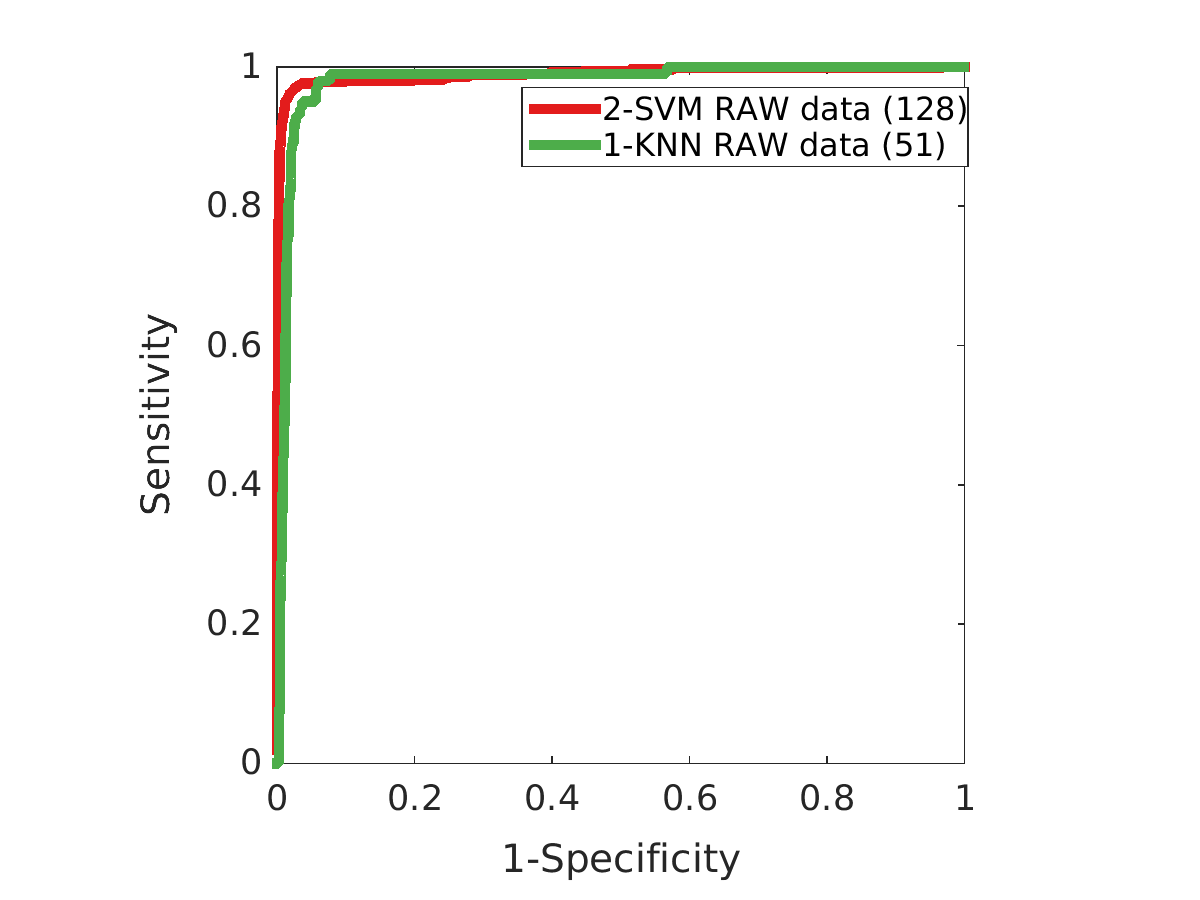} \\
  (a)&(b)
   \end{tabular}
  \caption{ROC curves corresponding to the comparison between the best novelty detector and the best two-classes detector. (a) Collection 1. (b) Collection 2. }
  \label{fig:roc}
\end{figure*}

The results achieved by the SVM classifier in the case of 128 samples make clear that the \emph{two-classes} classifier performs better than a novelty detector. This is not true in the case of 51 samples where we demonstrated that using raw data the gap between SVM and the novelty detector is very small. This results overcome the results achieved by~\citet{medrano2014}. They demonstrated that  a \emph{two-classes} SVM is much better than a novelty detector when the accelerometer data is represented as \emph{magnitude} and is composed of  51 samples. This result is more visible looking at the table \ref{table:summary}. This table includes the best results achieved by a two-classes and a one-class classifier. It is quite evident that the novelty detector, based on the kNN classifier, achieves a performance that is about 10\% less than the the two-classes SVM classifier. The ROC curves that compare the best one-class versus the best two-classes classifier performed on the collection 1 and 2 are showed in Fig.~\ref{fig:roc}. Also from these figures is quite evident that the performance of the novelty detector is very close to the one of the two-classes classifier. 

\begin{table}[tb]
\scriptsize
\centering
\begin{tabular}{llccccc}
C.&	Feat.&Class.	&	$AUC$ &	$SE$ &	$SP$	&	$\sqrt{SE\cdot SP}$ \\
\hline
\multirow{16}{*}{1}	&\multirow{3}{*}{RAW}	&1-knn&	0.955&	0.950&	0.899&	0.924\\ 
&&2-knn&	0.969&	0.932&	0.927&	0.929\\ 
&&1-svm&	0.939&	0.936&	0.867&	0.901\\ 
&&2-svm&	\bf{0.989}&	\bf{0.964}&	\bf{0.981}&	\bf{0.972}\\ 
[2pt] 
&\multirow{3}{*}{Magn.}	&1-knn&	0.822&	0.900&	0.656&	0.768\\ 
&&2-knn&	0.876&	0.834&	0.761&	0.794\\ 
&&1-svm&	0.765&	0.880&	0.615&	0.735\\ 
&&2-svm&	\bf{0.977}&	\bf{0.924}&	\bf{0.954}&	\bf{0.939}\\ 
[2pt] 
&\multirow{3}{*}{Energ.}	&1-knn&	0.811&	0.870&	0.695&	0.777\\ 
&&2-knn&	0.925&	0.910&	0.890&	0.900\\ 
&&1-svm&	0.843&	0.856&	0.768&	0.811\\ 
&&2-svm&	\bf{0.988}&	\bf{0.970}&	\bf{0.958}&	\bf{0.964}\\ 
[2pt] 
&\multirow{3}{*}{LTP}	&1-knn&	0.817&	0.850&	0.668&	0.754\\ 
&&2-knn&	0.853&	0.876&	0.697&	0.780\\ 
&&1-svm&	0.793&	0.846&	0.637&	0.734\\ 
&&2-svm&	\bf{0.974}&	\bf{0.910}&	\bf{0.940}&	\bf{0.925}\\ 
[2pt] 
\hline 
\multirow{16}{*}{2}	&\multirow{3}{*}{RAW}	&1-knn&	0.977&	0.960&	0.933&	0.947\\ 
&&2-knn&	0.984&	0.958&	0.947&	0.952\\ 
&&1-svm&	0.971&	0.942&	0.928&	0.935\\ 
&&2-svm&	\bf{0.992}&	\bf{0.976}&	\bf{0.986}&	\bf{0.981}\\ 
[2pt] 
&\multirow{3}{*}{Magn.}	&1-knn&	0.885&	0.910&	0.762&	0.833\\ 
&&2-knn&	0.928&	0.898&	0.814&	0.854\\ 
&&1-svm&	0.804&	0.908&	0.661&	0.775\\ 
&&2-svm&	\bf{0.984}&	\bf{0.944}&	\bf{0.957}&	\bf{0.950}\\ 
[2pt] 
&\multirow{3}{*}{Energ.}	&1-knn&	0.891&	0.870&	0.803&	0.836\\ 
&&2-knn&	0.916&	0.860&	0.878&	0.868\\ 
&&1-svm&	0.918&	0.910&	0.851&	0.880\\ 
&&2-svm&	\bf{0.990}&	\bf{0.968}&	\bf{0.983}&	\bf{0.975}\\ 
[2pt] 
&\multirow{3}{*}{LTP}	&1-knn&	0.885&	0.910&	0.754&	0.828\\ 
&&2-knn&	0.917&	\bf{0.930}&	0.779&	0.850\\ 
&&1-svm&	0.842&	0.898&	0.684&	0.784\\ 
&&2-svm&	\bf{0.984}&	0.926&	\bf{0.956}&	\bf{0.941}\\ 
[2pt] 
\hline 
\multirow{16}{*}{3}	&\multirow{3}{*}{RAW}	&1-knn&	0.998&	0.990&	0.984&	0.987\\ 
&&2-knn&	0.997&	0.996&	\bf{0.993}&	\bf{0.994}\\ 
&&1-svm&	0.998&	0.986&	0.987&	0.986\\ 
&&2-svm&	\bf{0.999}&	\bf{1.000}&	0.974&	0.987\\ 
[2pt] 
&\multirow{3}{*}{Magn.}	&1-knn&	0.991&	0.990&	0.972&	0.981\\ 
&&2-knn&	0.998&	0.996&	0.992&	0.994\\ 
&&1-svm&	0.637&	0.996&	0.600&	0.773\\ 
&&2-svm&	\bf{1.000}&	\bf{1.000}&	\bf{0.997}&	\bf{0.999}\\ 
[2pt] 
&\multirow{3}{*}{Energ.}	&1-knn&	0.898&	0.890&	0.767&	0.826\\ 
&&2-knn&	0.930&	0.856&	0.907&	0.880\\ 
&&1-svm&	0.996&	0.998&	0.974&	0.986\\ 
&&2-svm&	\bf{1.000}&	\bf{1.000}&	\bf{0.999}&	\bf{1.000}\\ 
[2pt] 
&\multirow{3}{*}{LTP}	&1-knn&	0.988&	0.970&	0.964&	0.967\\ 
&&2-knn&	0.997&	0.990&	0.980&	0.985\\ 
&&1-svm&	0.980&	0.944&	0.914&	0.929\\ 
&&2-svm&	\bf{1.000}&	\bf{0.996}&	\bf{0.999}&	\bf{0.997}\\ 
[2pt] 
\hline 
\end{tabular}
\medskip
\caption{Results obtained by both kNN and SVM schemas on the three collections. Here the accelerometer data contain 128 samples. Best result for each collection is reported in bold.}
\label{table:results1}
 \end{table}

\begin{table}[tb]
\scriptsize
\centering
\begin{tabular}{llccccc}
C.&	Feat.&Class.	&	$AUC$ &	$SE$ &	$SP$	&	$\sqrt{SE\cdot SP}$ \\
\hline
\multirow{16}{*}{1}	&\multirow{3}{*}{RAW}	&1-knn&	0.980&	\bf{0.980}&	0.940&	0.960\\ 
&&2-knn&	0.983&	0.962&	0.952&	0.957\\ 
&&1-svm&	0.954&	0.946&	0.903&	0.924\\ 
&&2-svm&	\bf{0.986}&	0.954&	\bf{0.969}&	\bf{0.961}\\ 
[2pt] 
&\multirow{3}{*}{Magn.}	&1-knn&	0.958&	\bf{0.910}&	0.923&	0.916\\ 
&&2-knn&	0.961&	0.910&	0.929&	0.919\\ 
&&1-svm&	0.911&	0.870&	0.844&	0.857\\ 
&&2-svm&	\bf{0.967}&	0.904&	\bf{0.948}&	\bf{0.926}\\ 
[2pt] 
&\multirow{3}{*}{Energ.}	&1-knn&	0.811&	0.870&	0.695&	0.777\\ 
&&2-knn&	0.925&	0.910&	0.890&	0.900\\ 
&&1-svm&	0.843&	0.856&	0.768&	0.811\\ 
&&2-svm&\bf{0.988}&	\bf{0.970}&	\bf{0.958}&	\bf{0.964}\\ 
[2pt] 
&\multirow{3}{*}{LTP}	&1-knn&	0.936&	0.860&	0.889&	0.875\\ 
&&2-knn&	0.942&	0.882&	0.891&	0.885\\ 
&&1-svm&	0.890&	0.826&	0.820&	0.823\\ 
&&2-svm &\bf{0.959}&	\bf{0.890}&	\bf{0.923}&	\bf{0.906}\\ 
[2pt] 
\hline 
\multirow{16}{*}{2}	&\multirow{3}{*}{RAW}	&1-knn&	0.988&	0.960&	0.978&	0.969\\ 
&&2-knn&	0.990&	0.964&	0.977&	0.970\\ 
&&1-svm&	0.979&	0.948&	0.954&	0.951\\ 
&&2-svm&\bf{0.990}&	\bf{0.960}&	\bf{0.981}&	\bf{0.970}\\ 
[2pt] 
&\multirow{3}{*}{Magn.}	&1-knn&	0.970&	0.940&	0.926&	0.933\\ 
&&2-knn&	0.976&	0.942&	0.938&	0.940\\ 
&&1-svm&	0.942&	0.894&	0.879&	0.887\\ 
&&2-svm&	\bf{0.984}&	\bf{0.952}&	\bf{0.955}&	\bf{0.953}\\ 
[2pt] 
&\multirow{3}{*}{Energ.}	&1-knn&	0.891&	0.870&	0.803&	0.836\\ 
&&2-knn&	0.916&	0.860&	0.878&	0.868\\ 
&&1-svm&	0.918&	0.910&	0.851&	0.880\\ 
&&2-svm&	\bf{0.990}&	\bf{0.972}&	\bf{0.988}&	\bf{0.980}\\ 
[2pt] 
&\multirow{3}{*}{LTP}	&1-knn&	0.958&	0.900&	0.902&	0.901\\ 
&&2-knn&	0.964&	\bf{0.938}&	0.894&	0.915\\ 
&&1-svm&	0.930&	0.900&	0.832&	0.865\\ 
&&2-svm&	\bf{0.980}&	0.934&	\bf{0.933}&	\bf{0.934}\\ 
[2pt] 
\hline 
\multirow{16}{*}{3}	&\multirow{3}{*}{RAW}	&1-knn&	0.998&	\bf{1.000}&	0.974&	0.987\\ 
&&2-knn&	0.997&	1.000&	0.995&	\bf{0.998}\\ 
&&1-svm&	0.999&	0.986&	\bf{0.997}&	0.992\\ 
&&2-svm&	\bf{1.000}&	0.996&	0.985&	0.991\\ 
[2pt] 
&\multirow{3}{*}{Magn.}	&1-knn&	0.996&	0.990&	\bf{0.993}&	0.991\\ 
&&2-knn&	0.998&	0.998&	0.991&	0.994\\ 
&&1-svm&	0.763&	0.998&	0.741&	0.860\\ 
&&2-svm&	\bf{0.999}&	\bf{0.998}&	0.991&	\bf{0.995}\\ 
[2pt] 
&\multirow{3}{*}{Energ.}	&1-knn&	0.898&	0.890&	0.767&	0.826\\ 
&&2-knn&	0.930&	0.856&	0.907&	0.880\\ 
&&1-svm&	0.996&	0.998&	0.974&	0.986\\ 
&&2-svm&	\bf{1.000}&	\bf{1.000}&	\bf{0.999}&	\bf{1.000}\\ 
[2pt] 
&\multirow{3}{*}{LTP}	&1-knn&	0.996&	0.960&	\bf{0.991}&	0.975\\ 
&&2-knn&	0.997&	0.986&	0.987&	0.987\\ 
&&1-svm&	0.997&	0.986&	0.978&	0.982\\ 
&&2-svm&	\bf{1.000}&	\bf{0.998}&	0.989&	\bf{0.994}\\ 
[2pt] 
\hline 
\end{tabular}
\medskip
\caption{Results obtained by both kNN and SVM schemas on the three collections. Here the accelerometer data contain 51 samples. Best result for each collection is reported in bold.}
\label{table:results2}
 \end{table}

 \begin{table}[tb]
\scriptsize
\centering
\begin{tabular}{lccccc}
	Features&Class.	&	$AUC$ &	$SE$ &	$SP$	&	$\sqrt{SE\cdot SP}$ \\
\hline
\multirow{4}{*}{RAW}	&1-knn&	0.977&	0.967&	0.939&	0.953\\ 
&2-knn&	0.984&	0.962&	0.955&	0.959\\ 
&1-svm&	0.969&	0.955&	0.927&	0.941\\ 
&2-svm&	0.993&	0.980&	0.980&	0.980\\ 
[2pt] 
\hline 
\multirow{4}{*}{Magn.}	&1-knn&	0.900&	0.933&	0.797&	0.861\\ 
&2-knn&	0.934&	0.909&	0.855&	0.881\\ 
&1-svm&	0.736&	0.928&	0.625&	0.761\\ 
&2-svm&	0.987&	0.956&	0.969&	0.963\\ 
[2pt] 
\hline 
\multirow{4}{*}{Energ.}	&1-knn&	0.867&	0.877&	0.755&	0.813\\ 
&2-knn&	0.924&	0.875&	0.892&	0.883\\ 
&1-svm&	0.919&	0.921&	0.864&	0.892\\ 
&2-svm&	0.993&	0.979&	0.980&	0.980\\ 
[2pt] 
\hline 
\multirow{4}{*}{LTP}	&1-knn&	0.897&	0.910&	0.795&	0.850\\ 
&2-knn&	0.922&	0.932&	0.819&	0.872\\ 
&1-svm&	0.872&	0.896&	0.745&	0.815\\ 
&2-svm&	0.986&	0.944&	0.965&	0.954\\ 
[2pt] 
\hline  
\end{tabular}
\medskip
\caption{Average results obtained by both kNN and SVM schemas with respect to the three collections. Here the accelerometer data contain 128 samples.}
\label{table:summary-1-128}
 \end{table}
 
   \begin{table}[tb]
\scriptsize
\centering
\begin{tabular}{lccccc}
	Features&Class.	&	$AUC$ &	$SE$ &	$SP$	&	$\sqrt{SE\cdot SP}$ \\
\hline
\multirow{4}{*}{RAW}	&1-knn&	0.989&	0.980&	0.964&	0.972\\ 
&2-knn&	0.990&	0.975&	0.975&	0.975\\ 
&1-svm&	0.977&	0.960&	0.952&	0.956\\ 
&2-svm&	0.992&	0.970&	0.978&	0.974\\ 
[2pt] 
\hline 
\multirow{4}{*}{Magn.}	&1-knn&	0.975&	0.947&	0.947&	0.947\\ 
&2-knn&	0.978&	0.950&	0.953&	0.951\\ 
&1-svm&	0.872&	0.921&	0.822&	0.868\\ 
&2-svm&	0.983&	0.951&	0.965&	0.958\\ 
[2pt] 
\hline 
\multirow{4}{*}{Energ.}	&1-knn&	0.867&	0.877&	0.755&	0.813\\ 
&2-knn&	0.924&	0.875&	0.892&	0.883\\ 
&1-svm&	0.919&	0.921&	0.864&	0.892\\ 
&2-svm&	0.993&	0.981&	0.982&	0.981\\ 
[2pt] 
\hline 
\multirow{4}{*}{LTP}	&1-knn&	0.964&	0.907&	0.927&	0.917\\ 
&2-knn&	0.968&	0.935&	0.924&	0.929\\ 
&1-svm&	0.939&	0.904&	0.877&	0.890\\ 
&2-svm&	0.979&	0.941&	0.948&	0.944\\ 
[2pt] 
\hline 
\end{tabular}
\medskip
\caption{Average results obtained by both kNN and SVM schemas with respect to the three collections. Here the accelerometer data contain 51 samples.}
\label{table:summary-1-51}
 \end{table}

 \begin{table}[tb]
\scriptsize
\centering
\begin{tabular}{lccccc}
	Features&Collect.	&	$AUC$ &	$SE$ &	$SP$	&	$\sqrt{SE\cdot SP}$ \\
\hline
\multirow{4}{*}{RAW}	&1&	0.963&	0.946&	0.918&	0.932\\ 
&2&	0.981&	0.959&	0.949&	0.954\\ 
&3&	0.998&	0.993&	0.984&	0.989\\ 
[2pt] 
\hline 
\multirow{4}{*}{Magn.}	&1&	0.860&	0.884&	0.746&	0.809\\ 
&2&	0.900&	0.915&	0.799&	0.853\\ 
&3&	0.906&	0.996&	0.890&	0.937\\ 
[2pt] 
\hline 
\multirow{4}{*}{Energ.}	&1&	0.892&	0.901&	0.828&	0.863\\ 
&2&	0.929&	0.902&	0.879&	0.890\\ 
&3&	0.956&	0.936&	0.912&	0.923\\ 
[2pt] 
\hline 
\multirow{4}{*}{LTP}	&1&	0.859&	0.871&	0.736&	0.798\\ 
&2&	0.907&	0.916&	0.793&	0.851\\ 
&3&	0.991&	0.975&	0.964&	0.969\\ 
[2pt] 
\hline 
\end{tabular}
\medskip
\caption{Average results obtained by both kNN and SVM schemas with respect to the four data representation. Here the accelerometer data contain 128 samples.}
\label{table:summary-2-128}
 \end{table}

 \begin{table}[tb]
\scriptsize
\centering
\begin{tabular}{lccccc}
	Features&Collect.	&	$AUC$ &	$SE$ &	$SP$	&	$\sqrt{SE\cdot SP}$ \\
\hline
\multirow{4}{*}{RAW}	&1&	0.976&	0.960&	0.941&	0.951\\ 
&2&	0.987&	0.958&	0.973&	0.965\\ 
&3&	0.998&	0.995&	0.988&	0.992\\ 
[2pt] 
\hline 
\multirow{4}{*}{Magn.}	&1&	0.949&	0.898&	0.911&	0.905\\ 
&2&	0.968&	0.932&	0.924&	0.928\\ 
&3&	0.939&	0.996&	0.929&	0.960\\ 
[2pt] 
\hline 
\multirow{4}{*}{Energ.}	&1&	0.892&	0.901&	0.828&	0.863\\ 
&2&	0.929&	0.903&	0.880&	0.891\\ 
&3&	0.956&	0.936&	0.912&	0.923\\ 
[2pt] 
\hline 
\multirow{4}{*}{LTP}	&1&	0.932&	0.865&	0.881&	0.872\\ 
&2&	0.958&	0.918&	0.890&	0.904\\ 
&3&	0.998&	0.982&	0.986&	0.984\\ 
[2pt] 
\hline 
\end{tabular}
\medskip
\caption{Average results obtained by both kNN and SVM schemas with respect to the four data representation. Here the accelerometer data contain 51 samples.}
\label{table:summary-2-51}
 \end{table}
 
  \begin{table}[tb]
\scriptsize
\centering
\begin{tabular}{lcccc}
	Class.&	$AUC$ &	$SE$ &	$SP$	&	$\sqrt{SE\cdot SP}$ \\
\hline
1-knn&	0.910&	0.922&	0.822&	0.869\\ 
[2pt] 
2-knn&	0.941&	0.920&	0.880&	0.898\\ 
[2pt] 
1-svm&	0.874&	0.925&	0.790&	0.852\\ 
[2pt] 
2-svm&	0.990&	0.965&	0.974&	0.969\\ 
[2pt] 
\hline 
\end{tabular}
\medskip
\caption{Average results obtained by both all classifiers with respect to the four data representation and the three collections. Here the accelerometer data contain 128 samples.}
\label{table:summary-3-128}
 \end{table}

  \begin{table}[tb]
\scriptsize
\centering
\begin{tabular}{lcccc}
	Class.&	$AUC$ &	$SE$ &	$SP$	&	$\sqrt{SE\cdot SP}$ \\
\hline
1-knn&	0.948&	0.927&	0.898&	0.912\\ 
[2pt] 
2-knn&	0.965&	0.934&	0.936&	0.934\\ 
[2pt] 
1-svm&	0.927&	0.926&	0.879&	0.901\\ 
[2pt] 
2-svm&	0.987&	0.961&	0.968&	0.964\\ 
[2pt] 
\hline 
\end{tabular}
\medskip
\caption{Average results obtained by both all classifiers with respect to the four data representation and the three collections. Here the accelerometer data contain 51 samples.}
\label{table:summary-3-51}
 \end{table}

 \begin{table}[t]
\scriptsize
\centering
\begin{tabular}{llccccc}
	Coll.&Features&Class.	&	$AUC$ &	$SE$ &	$SP$	&	$\sqrt{SE\cdot SP}$ \\
\hline
\multirow{2}{*}{1}	&RAW(128)	&2-svm&0.989&	0.964&	0.981&	0.972\\ 
&RAW(51)	&1-knn&	0.980&0.980&	0.940&	0.960\\  
[2pt]  
 \hline
\multirow{2}{*}{2}	&RAW(128)	&2-svm&	0.992&	0.976&	0.986&	0.981\\ 
&RAW(51)	&1-knn&	0.988&	0.960&	0.978&	0.969\\ 
[2pt]  
 \hline
\multirow{2}{*}{3}	&Energ.(128)	&2-svm&	1.000&	1.000&	0.999&	1.000\\ 
&RAW(51)	&1-svm&	0.999&	0.986&	0.997&	0.992\\ 
[2pt] 
\hline
\end{tabular}
\medskip
\caption{Results obtained by the best feature vectors in the case of both the two-classes and one-class classifiers.}
\label{table:summary}
 \end{table}

\section{Discussion and Conclusion}
\label{sec:discussion}

In this work we evaluated the robustness of anomaly detectors (\emph{one-class} classifier) compared to that of traditional \emph{two-classes} detection methods that, in turn, are tuned with fall instances. To this end, we experimented  several methods  on three different collections of accelerometer data, and four different feature vectors. The experiments have demonstrated that:
\begin{itemize}
\item a very simple feature vector based on raw data is very robust to detect falls in both one or two classes schemas;
\item a greater number of samples of acceleration instances penalises kNN classification schemas. In contrast, the SVM classifier does not seem to suffer from changes in the number of samples. This makes the one-class kNN classifier more feasible in case of  51 samples;
\item  in the case of 128 samples a novelty detector is reliable only if it is based on raw data. In the case of 51 samples a novelty detector is reliable if it is based on both raw data and magnitude;
\end{itemize}

Overall, considering that in the case of raw data, the gap between the SVM and one class kNN is very small, we can conclude that a fall detection system based on a novelty detector is feasible in a real scenario. This is especially true considering the limited computation capacity and power resources of the smartphone. In fact, the raw data does not require further processing and the kNN schema is based on a simple Euclidean distance. 

\section{Future Directions}
\label{sec:future}
In order to further validate the robustness of our approach, we should be able to experiment with additional datasets. These datasets should contain ADL performed by different people and recorded by different smartphones.

As the number of data sets freely available is extremely reduced, we decided to develop an application that is able to acquire data from smartphones' sensors and to automatically label them (falls or ADL). This enables us to enrich the datasets of ADL and of simulated falls.


\begin{acknowledgements}
We would like to thank the Reviewers for their valuable comments and suggestions that allowed us to improve the paper.
\end{acknowledgements}


\end{document}